\documentclass[twocolumn,showpacs,preprintnumbers,amsmath,amssymb,floatfix]{revtex4}
\usepackage{graphicx}

\begin{document}

\title{Quantum coherence in a ferromagnetic metal: time-dependent conductance fluctuations}

\author{Sungbae Lee, Aaron Trionfi, and Doug Natelson}

\affiliation{Department of Physics and Astronomy, Rice University, 6100 Main St., Houston, TX 77005}

\date{\today}


\begin{abstract}
Quantum coherence of electrons in ferromagnetic metals is difficult to
assess experimentally.  We report the first measurements of
time-dependent universal conductance fluctuations in ferromagnetic
metal (Ni$_{0.8}$Fe$_{0.2}$) nanostructures as a function of
temperature and magnetic field strength and orientation.  We find that
the cooperon contribution to this quantum correction is suppressed,
and that domain wall motion can be a source of coherence-enhanced
conductance fluctuations.  The fluctuations are more strongly
temperature dependent than those in normal metals, hinting that an
unusual dephasing mechanism may be at work.
\end{abstract}

\maketitle

While quantum coherence effects in normal
metals are well established\cite{WashburnetAl92RPP,Imry97book}, few
experimental examinations of such physics have been performed in
ferromagnetic (FM) metals.  Quantum corrections to conduction in FM
systems are of fundamental interest due to correlation-induced degrees
of freedom not present in normal metals ({\it e.g.} spin waves), and
the interplay of FM order with coherence.  These corrections have been
discussed while considering\cite{TataraetAl97PRL} domain walls effects
on conduction in FM nanowires\cite{HongetAl95PRB}, and
magnetoresistive effects\cite{DugaevetAl01PRB} in thin FM
films\cite{AprilietAl97SSC}.  The conductance of a mesoscopic
ferromagnet is expected\cite{TataraetAl97PRL} to be highly sensitive
to domain wall motion, just as the conductance of a mesoscale normal
metal is sensitive to the motion of an individual elastic
scatterer\cite{AltshuleretAl85JETPL,FengetAl86PRL}.  Domain wall
motion can thus lead to time-dependent (TD) universal conductance
fluctuations (UCF)\cite{LeeetAl85PRL,LeeetAl87PRB}.  Further
nontrivial coherence effects proposed in FMs include Berry's phase
physics due to coherent diffusion of spins through nonuniform
magnetization\cite{LyandaGelleretAl98PRL}, and dephasing via domain
wall\cite{TataraetAl97PRL,LyandaGelleretAl98PRL,KoyamaetAl03JPSJ} and
spin wave\cite{Takane03JPSJ} scattering.  Quantum coherence in FM
systems may also have technological relevance in understanding novel
mesoscale devices based on FM semiconductors.

No systematic experimental examination has been reported of electronic
coherence in FM materials and the effects of domains.  Two common
techniques for assessing quantum coherence in normal metals, weak
localization (WL)\cite{Bergmann84PR} and UCF as a function of field,
rely on magnetoresistive measurements.  In FM metals the anisotropic
magnetoresistance (AMR), a bandstructure effect, significantly
complicates efforts to find WL and UCF.  Examinations of mesoscale
magnetic structures hint at UCF\cite{AumentadoetAl00PB}, and
Aharanov-Bohm (AB) oscillations have been observed in multidomain and
single-domain mesoscopic NiFe
rings\cite{KasaietAl03JAP,SaitohetAl03PE} at mK temperatures.

TDUCF noise is an alternative probe of electronic
coherence\cite{BirgeetAl89PRL,Giordanobook}, and noise measurements
are comparatively immune to the obscuring effects of AMR.  TDUCF noise
results from the motion of localized defects, thought to be tunneling
two-level systems (TLS)\cite{Esquinazibook} in the polycrystalline
metal.  With the typical TLS relaxation time distribution, the
resulting resistance noise power, $S_{R}$, has a $1/f$
dependence\cite{FengetAl86PRL}.  In a nonmagnetic metal, applied
magnetic flux suppresses the cooperon contribution to the
fluctuations\cite{Stone89PRB} relative to the diffuson over a field
scale related to the coherence length, $L_{\phi}^{\rm TDUCF}$,
reducing $S_{R}$ by a factor of two.  In spin glasses TDUCF noise
resulting from slow rearrangements of frozen spin configurations has
also been observed\cite{IsraeloffetAl89PRL,NeuttiensetAl00PRB}.

We report the first TDUCF measurements on a FM metal, permalloy
(Ni$_{0.8}$Fe$_{0.2}$).  There is no discernable difference between
$S_{R}(T)$ at zero external magnetic field and large external fields
(several Tesla), indicating suppression of the cooperon contribution
to the TDUCF due to the FM state.  Additional noise is found over a
range of low temperatures and fields when domain walls are favored,
demonstrating experimentally that domain walls can act as coherent
scatterers of carriers.  We argue that the fluctuators responsible for
TDUCF in disordered FM metals are likely TLS similar to those in
disordered normal metals.  If this is so, the unusual $S_{R}(T)\sim
T^{-2}$ as $T\rightarrow 0$ may suggest an unconventional dephasing
mechanism in FM materials.

\begin{figure}[h!]
\begin{center}
\includegraphics[clip, width=7.5cm]{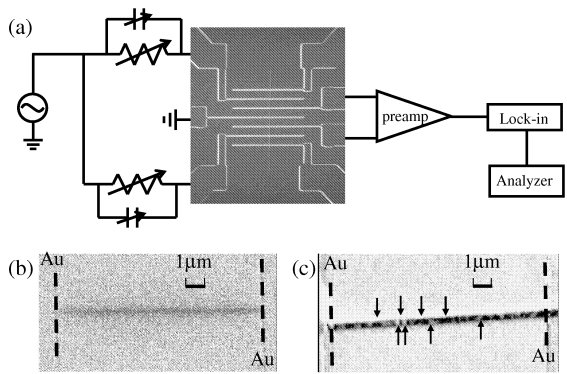}
\end{center}
\vspace{-3mm}
\caption{(a) Noise measurement scheme. Only five 
consecutive leads are used for the noise measurement among seven leads. 
(b) MFM image of one segment of 27~nm wide wire. The whole segment is a 
single domain. Left and right edges of image are Ti/Au leads covering the wire.  
(c) MFM image for 450~nm wide wire. Arrows indicate some of the domain walls. }
\label{fig:config}
\vspace{-3mm}
\end{figure}

Each sample is fabricated by two steps of electron beam lithography on
an undoped GaAs substrate.  A permalloy (Py) wire $t=$10~nm thick is
patterned and deposited by electron beam evaporation from a
Ni$_{0.8}$Fe$_{0.2}$ source.  To minimize Py oxidation, e-beam resist
is baked in a forming gas environment.  Current and voltage leads are
then patterned, and the exposed contact areas are ion-milled for 40
seconds immediately before evaporation of leads.  Contact resistances
are $< 100~\Omega$, a factor of five lower than in samples without
ion-milling.  The leads are 1.5~nm Ti / 40~nm Au, 1~$\mu$m wide.  Each
segment of wire between leads is 10 $\mu$m in length.
Figure~\ref{fig:config}(a) shows the sample configuration, and the
sample parameters are given in Table~\ref{tab:samples}.

\begin{table}
\caption{Samples used in magnetotransport and noise measurements. All samples
are 10~nm thick permalloy, and each segment is 10~$\mu$m in length.}
\begin{tabular}{c c c c c}
\hline \hline
Sample & $w$~[nm] & $\rho$(T=2~K)~[$\mu\Omega$-cm] & AMR~(2~K) & $B_{\rm sw}$ [T]  \\
\hline
A & 27 & 44.86 &  3.1\% & 0.47  \\
B & 50 & 48.63 &  3.5\% & 0.63 \\
C & 100 & 50.87 & 3.8\% & 0.59 \\
D & 450 & 50.55 & 3.7\% & 0.34 \\
\hline
\hline
\end{tabular}
\label{tab:samples}
\vspace{-3mm}
\end{table}

Since permalloy is a relatively soft ferromagnet, geometric anisotropy
strongly influences domain configurations.  Increasing wire aspect
ratio favors a single-domain configuration with magnetization along
the wire axis.  Nonmagnetic leads preserve a simple FM geometry and
maintain uniaxial geometric anisotropy.  Room temperature domain
structures are investigated by magnetic force microscopy
(MFM)\cite{NozakietAl99JJAP}.  Figures~\ref{fig:config}(b) and (c) are
MFM images for 27 and 450~nm wide wires respectively, after exposure
to applied fields perpendicular to the wires.  With no external
fields, 100 and 450~nm wide wires show multiple domain walls within
one segment, whereas long single domain features are observed for the
smaller wires, with few domain walls throughout all six segments.

Samples are measured in a $^{4}$He cryostat, with magnetic fields
either along the wire axis (``parallel'') or normal to the plane of
the wire (``perpendicular'').  Sample resistance as a function of
temperature is measured with standard four-terminal techniques, and
Joule heating effects are characterized via $R(T)$.  Noise
measurements use a five-terminal ac bridge
technique(Fig.~\ref{fig:config}(a)) \cite{Scofield87RSI,TrionfietAl04}
at $\sim$600~Hz, with the demodulated signal examined between 78~mHz
and a few Hz.  The high currents necessary for the noise measurements
limit these measurements to 2~K and higher.  No significant drive
current or frequency dependence was observed in either
magnetoresistance or $S_{R}$ until currents were large enough to
affect $R(T)$.

\begin{figure}[h!]
\begin{center}
\includegraphics[clip, width=7.5cm]{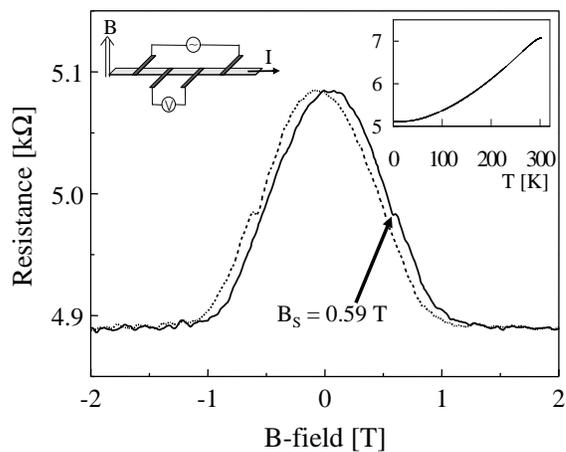}
\end{center}
\vspace{-6mm}
\caption{
Anisotropic magnetoresistance data with a perpendicular magnetic field
for the 100~nm wide wire at 8~K. Resistance of the sample was recorded
while sweeping magnetic field between $\pm$2~T. Solid line is for
sweeping the field from -2~T to +2~T and dashed line is for sweeping
the other direction.  Arrow indicates the discontinuity due to
magnetization reorientation, and the switching field, $B_{S}$, for
this sample was $\sim$0.59~T.  Left inset shows four terminal scheme
used for the AMR measurement. Right inset shows $R$ vs. $T$ for this
sample (slight rise below 10~K not apparent on this scale).}
\label{fig:amr}
\vspace{-3mm}
\end{figure}

Figure~\ref{fig:amr} shows typical AMR data in the 100~nm wide wire at
8~K, with $B_{\perp}$ sweeping at 100~Oe/s between $\pm$~2~T.  The
data are independent of sweep rate below a cutoff set by the data
acquisition speed.  At low fields the geometric anisotropy aligns the
magnetization, ${\bf M}$, with the wire axis and hence the current
density ${\bf J}$.  At large values of $B_{\perp}$, ${\bf M}\cdot{\bf
J} = 0$, leading to a lower resistivity via the AMR.  Analogous
measurements with $B_{||}$ show essentially no AMR signal, consistent
with full alignment of ${\bf M}$ along the wire at zero field.  The
AMR resistivity ratio $(R_{||}-R_{\perp})/R_{||}$ of 3-4~\% agrees
with previous results \cite{McGuireetAl75IEEE,RijksetAl97PRB}.  The MR
discontinuity occurs at the switching field, $B_{\rm s}$,
corresponding to reorienting unstable domains\cite{WegroweetAl99PRL}.
The switching field decreases with increasing temperature.  When
$B_{\perp}>>B_{\rm s}$, the magnetization, ${\bf M}$, is believed to
be uniform and aligned with the applied external field.  Disorder and
edge effects may cause the local magnetization to deviate from the
bulk value\cite{AumentadoetAl99APL}.  No detectable WL
magnetoresistance was seen in any of the samples, down to 1.7~K.

At low temperatures, the voltage noise power is very well
described by a $1/f$ dependence, and scales with the square of the drive
current, indicating that its source is a fluctuating sample
resistance.  The frequency dependence remains $1/f$ for for all
samples and all parallel and perpendicular fields examined between 0
and 8~T, even when the magnetic field is very close to $B_{\rm s}$.

\begin{figure}[h!]
\begin{center}
\includegraphics[clip, width=7.5cm]{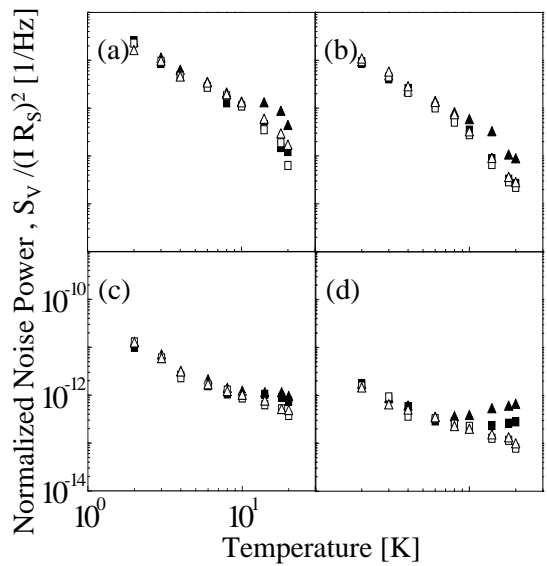}
\end{center}
\vspace{-5mm}
\caption{Noise power as a function of temperature for all four samples: 
(a) $w=$27~nm, (b) 50~nm, (c) 100~nm, and (d) 450~nm. For all four sets 
of data, solid squares are B=0~T; open squares are $B_{\perp}=8$~T;  open 
triangles are $B_{||}=$8~T; and solid triangle is $B_{\perp} \sim B_{\rm s}$.  
Error bars are comparable in size to the symbols.}
\label{fig:noiseT}
\vspace{-5mm}
\end{figure}


Figure~\ref{fig:noiseT} shows $S_{R}(T)$ for the permalloy wire
samples under a variety of conditions.  That this noise is TDUCF is
supported by several facts.  First, the noise power increases as
temperature is decreased for $T < 50$~K; this is expected in TDUCF due
to reduced ensemble averaging as $k_{\rm B}T$ decreases relative to
the Thouless energy $\hbar D/L_{\phi}^{2}$, and $L_{\phi}(T)$ grows
relative to the sample size, $L$.  Second, the magnitude of the noise
power increases with decreasing sample cross-section, as is routinely
observed in TDUCF measurements of normal metals.  Furthermore, the
magnitude of the noise is comparable to that observed in TDUCF
measurements on a normal metal alloy, Au$_{0.6}$Pd$_{0.4}$ known to
have a short coherence length.

Quantitative analysis of TDUCF in normal metals is done via the
magnetic field dependent suppression of the cooperon contribution to
the noise power, mentioned above\cite{Stone89PRB}.  In these permalloy
wires, {\it no such decrease} in $S_{R}$ is observed.  As demonstrated
in Fig.~\ref{fig:noiseT}, $S_{R}(T, B=0)$ and $S_{R}(T, B_{\perp}=
8~{\rm T})$ are indistinguishable at low $T$.  While some
(diffuson-based) quantum coherence effects are observable in FM
nanostructures, this data and the lack of detectable WL suggest that
cooperon phenomena are suppressed in this material.  This also
implies, unfortunately, that the field-dependence of the TDUCF cannot
be used to analyze coherence lengths quantitatively in such FM metals.

The noise power data taken for $B_{\perp} \sim B_{\rm s}$ are
particularly interesting.  Noise measurements are not possible
precisely at $B_{\perp} = B_{\rm s}$ due to large, irreversible
fluctuations in bridge signal from domain rearrangements.  Instead,
noise data are acquired at fields 0.05-0.1~T away from $B_{\rm s}$
while sweeping $B \rightarrow B_{\rm s}$.  At these values of field
and between 10~K and 20~K, $S_{R}$ is enhanced relative to the
single-domain case.  Interestingly, the power spectrum of the noise
remains $1/f$ under these circumstances, implying the fluctuators
responsible for the noise continue to have a broad distribution of
relaxation times.

\begin{figure}[h!]
\begin{center}
\includegraphics[clip, width=7.5cm]{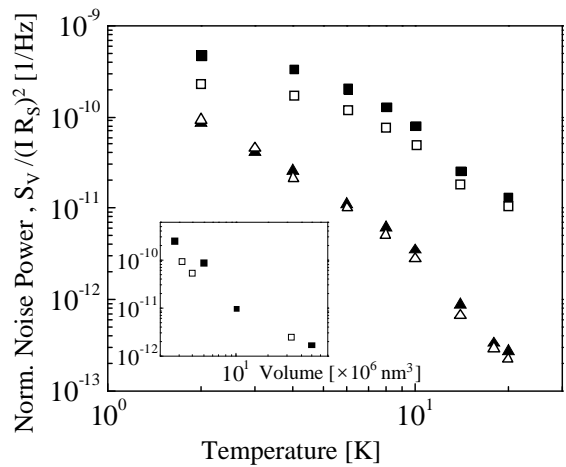}
\end{center}
\vspace{-3mm}
\caption{
Comparison of noise power plot between 50~nm wide Py wire (triangle
 points) and 35~nm wide nonmagnetic samples (square points). Solid
 symbols are for B=0~T and open symbols are for $B$=0.512~T for
 nonmagnetic wire and B=8~T for magnetic wire.  Nonmagnetic wire data 
 was shifted upward by a factor of 5 for clarity.  Inset:
Noise power as a function of sample volume at 2~K. Solid
 squares are for Py wires and open square are for nonmagnetic AuPd wires (identical sample geometry, with thicknesses from 6.5~nm to 9~nm).}
\label{fig:tdep}
\vspace{-2mm}
\end{figure}

One can consider whether the noise in this field regime is purely a
result of magnetization reorientation and AMR ({\it i.e.} not related
to electronic quantum coherence).  This is unlikely, since simple AMR
should be visible over all temperatures.  Rather, these data support
the idea\cite{TataraetAl97PRL} that domain walls can act as coherent
scatterers of electrons, and their motion can be a source of TDUCF.
The limited temperature range over which these effects are measurable
would then be determined by a combination of domain wall dynamics
(magnetization must fluctuate on the slow timescales probed by the
noise measurement) and the requirement of quantum coherence (as $T$
increases, TDUCF are increasingly suppressed because of thermal
averaging and decreasing $L_{\phi}(T)$).  More direct tests of these
ideas should be possible, {\it e.g.} in structures engineered to
contain only a single domain wall\cite{MiyakeetAl02JAP}.

Even though the cooperon contribution to TDUCF is apparently
suppressed in these FM samples, decoherence processes may still be
examined via the temperature dependence of $S_{R}$.
Figure~\ref{fig:tdep} compares the temperature dependence of the TDUCF
noise power in the 50~nm-wide Py sample and a 35~nm-wide nonmagnetic
Au$_{0.6}$Pd$_{0.4}$ wire.  The difference between low and high fields
is clear in the normal metal at low temperatures, while immeasurably
small in the ferromagnet, again demonstrating the suppression of the
cooperon contribution.  Note that $S_{R}(T) \sim T^{-2}$ between 2 and
8K for narrower permalloy wires.  While the temperature dependence is
slightly weaker in wider wires ($\sim T^{-1.4}$ for 450~nm wide wire),
it is {\it always} appreciably steeper that $1/T$.  This differs
significantly from the normal metal case, and may have implications for
dephasing mechanisms, as we discuss below.

With some assumptions, $S_{\rm R}(T)$ can be related to the coherence
length, $L_{\phi}(T)\equiv \sqrt{D \tau_{\phi}(T)}$, where $D$ is the
electronic diffusion constant and $\tau_{\phi}^{-1}$ is the
decoherence rate.  In normal metals, the predicted $S_{\rm R}(T)$
depends on the relationship between several length scales: segment
length $L$, width $w$, thickness $t$; the thermal length $L_{T} \equiv
\sqrt{\hbar D/ k_{\rm B}T}$; and $L_{\phi}(T)$.  Also relevant is the
density of active fluctuators, $n(T)$, in the material.  In the usual
TLS model, $n(T) \propto T$, and such TLS are apparent from the
low-$T$ acoustic properties of polycrystalline metals (see
Ref.~\cite{Esquinazibook}, Ch. 4).  In a quasi-2d system ($L,w >>
L_{\phi}, L_{T} >> t$), the expected temperature
dependence\cite{FengetAl86PRL,BirgeetAl89PRL} is $S_{R} \sim
n(T)L_{\rm min}^{2}L_{\phi}^{2}$, where $L_{\rm min}$ is the smaller
of $L_{T}$ and $L_{\phi}$.  In quasi-2d
bismuth\cite{BirgeetAl89PRL,BirgeetAl90PRB}, this TLS-based analysis
leads to an inferred $L_{\phi}\sim T^{-1/2}$, exactly as expected for
electron-electron dephasing in 2d.  For the normal metal data
shown\cite{TrionfietAl04}, the observed saturation of $S_{R}$ at low
temperatures is consistent with weak localization and TDUCF
measurements\cite{TrionfietAl04} on this sample that indicate the
presence of magnetic impurities in the AuPd and a resulting
$L_{\phi}(T)$ saturation.

We argue that the relevant fluctuators in the FM away from the
switching field are likely the same TLS as in normal metals (and hence
would have the same $n(T)$).  The lack of field dependence or
field-cooling effects show that the dominant fluctuators are {\it not}
moving domain walls or glassy
spins\cite{IsraeloffetAl89PRL,NeuttiensetAl00PRB}.  Furthermore, the
inset to Fig.~\ref{fig:tdep} shows that $S_{R}$ scales with sample
volume identically in normal metals (AuPd) and the permalloy.
Finally, the $1/f$ dependence of $S_{R}$ is identical between AuPd and
Py, showing that the fluctuators in both material systems have
identical distributions of relaxation times.  While not definitive,
these observations and the ubiquitous presence of TLS in a variety of
polycrystalline metals suggest that it is reasonable that such
fluctuators are active in the FM materials.

The dimensionality of our wires with respect to coherence phenomena 
remains unclear, though the most likely dimensionality is quasi-2d.  A
reasonable estimate for $D$ in these wires suggests that $L_{T} \sim
10$~nm at 10~K.  AB measurements in Py
rings\cite{KasaietAl03JAP,SaitohetAl03PE} at 30~mK estimate $L_{\phi}
\sim$ hundreds of nanometers, and is {\it much} shorter at 4.2~K.
$L_{\phi}$ values between 2~K and 20~K shorter than sample thinknesses
or much larger than sample widths seem incompatible with these AB
observations.

If $n(T) \sim T$ in these FM samples as in normal metals, then the
implications for $L_{\phi}(T)$ are interesting.  The unusually steep
$S_{R}\sim T^{-2}$ is stronger than that expected in {\it any}
dimensionality assuming standard electron-electron dephasing.
Furthermore, the identical temperature dependences of $S_{R}$ for low
and high $B_{||}$ are inconsistent with spin wave
scattering\cite{Takane03JPSJ} as the dominant decoherence mechanism,
since high $B_{||}$ is expected to exponentially suppress that
mechanism.  Lower temperature measurements should be revealing, as
would measurements independently assessing TLS properties in these FM
materials.

We present the first measurements of time-dependent universal
conductance fluctuations in ferromagnetic wires.  We find that the
cooperon contribution to the UCF is suppressed in this material.  We
also find evidence that domain wall motion leads to {\it enhanced}
conductance fluctuations, supporting the
statement\cite{TataraetAl97PRL} that domain walls may act as coherent
scatterers of carriers.  In single-domain configurations, the field
and sample size dependence of the noise power suggest that the
dominant fluctuators may be TLS similar to those in disordered normal
metals.  Within this picture, the steep $T$ dependence of the noise
power in the FM samples is surprising if conventional dephasing is at
work.  In concert with other techniques such as Aharanov-Bohm
measurements, it should be possible to examine decoherence in FM
systems quantitatively, and search for other novel coherence
effects\cite{Stern92PRL,LyandaGelleretAl98PRL}.

We thank I.L. Aleiner, B.L. Altshuler, N.O. Birge, and Q.M. Si for valuable 
conversations.  This work was supported by DOE grant
DE-FG03-01ER45946/A001.




\end{document}